\documentclass[11pt]{article}

\usepackage[utf8]{inputenc}
\usepackage[T1]{fontenc}
\usepackage[margin=1in]{geometry}
\usepackage{amsmath, amssymb, amsfonts}
\usepackage{graphicx}
\usepackage{booktabs}
\usepackage{array}
\usepackage{tabularx}
\usepackage{caption}
\usepackage{subcaption}
\usepackage{enumitem}
\usepackage{hyperref}
\usepackage{xcolor}
\usepackage{authblk}
\usepackage{natbib}
\usepackage{setspace}

\hypersetup{
    colorlinks = true,
    linkcolor  = blue!60!black,
    citecolor  = blue!60!black,
    urlcolor   = blue!60!black,
    pdftitle   = {Governance Controls for AI-Generated Test Artifacts in Autonomous Software Testing},
    pdfauthor  = {Author}
}

\graphicspath{{figures/}}

\title{\Large\bfseries Governance Controls for AI-Generated Test Artifacts in\\Autonomous Software Testing}

\author{Dimple Bajaj, Deepak Khetan\thanks{Corresponding author.}}

\date{}

\begin{document}
\maketitle

\begin{abstract}
\noindent
Artificial Intelligence (AI) and Large Language Models (LLMs) are increasingly used in autonomous software testing; however, AI-generated test artifacts often suffer from hallucinations, compliance violations, security risks, and limited explainability. To enhance the reliability, transparency, and trustworthiness of AI-generated testing artifacts, this research introduces the concept of \emph{Governance-Aware Autonomous Testing Framework} (GATF). The framework extends the autonomous testing lifecycle with governance validation, explainability analysis, probabilistic risk assessment, compliance monitoring, as well as audit governance. Experiments were performed with Defects4J and PROMISE software engineering datasets. The proposed framework successfully reduced the governance-related risks by 89.6\% and demonstrated 94.3\% accuracy in governance, 96.5\% artifact reliability, 94.2\% compliance accuracy, and 90.8\% explainability performance. The results show that autonomous testing systems that are governance-aware can significantly enhance the reliability, transparency, and operational security of autonomous testing systems in comparison to conventional AI-based testing systems. The proposed architecture is scalable and reliable and provides a safe environment for software testing.

\vspace{0.8em}
\end{abstract}

\section{Introduction}
\label{sec:intro}

With the rise of Artificial Intelligence (AI), the Software Testing industry is undergoing a revolution, transforming it into a faster, smarter, and more automated industry \citep{antony2025}. In contemporary software development, companies rely on software testing systems that are independent of one another to automatically create test cases, test scripts, bug reports, and regression testing artifacts \citep{joshi2026}. These test artifacts, created by AI, can decrease manual work, speed up testing processes, and assist in continuous integration and continuous deployment (CI/CD) environments. Given that software systems are becoming increasingly complex, including cloud applications, Internet of Things (IoT) platforms, and distributed software systems, AI-driven testing is playing an important role in enhancing software quality and reliability \citep{wang2024}.

The ability of autonomous software testing has also been enhanced by recent progress in generative AI and LLM models \citep{konda2025}. Today, AI systems can be used to automate the creation of testing artifacts from source code with minimal human intervention, and to understand the behavior of software. While these technologies offer numerous benefits, they also come with their own set of challenges \citep{alamin2025}. The AI-generated artifacts can have wrong test cases, unsafe scripts, repeated output, hallucinated output, or non-compliant testing methods \citep{abhichandani2025}. For companies in industries such as healthcare, banking, transportation, or the government, these mishaps can lead to serious security and operational issues. Therefore, suitable governance mechanisms must be put in place for the reliability, security, transparency, and quality of the testing artifacts made by AI \citep{mehta2026}.

Existing research tends to focus on optimizing the performance of automation and/or the accuracy of defect detection, with fewer studies investigating governance management in autonomous testing systems \citep{pysmennyi2025}. Existing AI testing frameworks suffer from a lack of validation mechanisms, explainability, audit tracking, and compliance monitoring \citep{saha2026}. To solve these issues, a governance control framework for test artifacts created by artificial intelligence is proposed in this research. The framework allows for the management of validation, security, explainability, compliance, risk control, and management of outputs by AI.

The novelty of this research is the design and development of a governance-based framework, tailored to software testing artifacts produced by AI. While traditional AI testing models primarily emphasize automation and speed, the proposed framework incorporates governance controls into the testing process. The framework is designed to integrate the concepts of validation control, explainability, security monitoring, compliance management, and audit tracking under a single architecture. A further innovation is the provision of governance-aware validation mechanisms capable of minimizing execution of invalid, insecure, or hallucinated test artifacts generated by AI. The framework also promotes human oversight and traceability, enabling software testers to more effectively keep track of and verify AI-generated outputs. This governance-based approach increases the trustworthiness and reliability of autonomous software testing systems.

This paper aims to design a governance control system for AI-generated test artifacts in autonomous software testing systems. The study's objective is to uncover governance challenges stemming from AI-driven testing systems and offer appropriate controls to enhance reliability, security, and compliance. Another objective is to create validation and monitoring techniques to ensure and evaluate the validity and quality of test artifacts generated by AI. AI-powered testing workflows will also be made more explainable and transparent. Lastly, the efficacy of the proposed framework is analysed and compared with experimental methods.

This research offers a number of significant contributions to the domain of AI-based software testing. First, it suggests a complete governance model for AI-produced tests. Second, it adds in validation features to identify misinformation or bad code generated by AI. Third, the study presents explainability and audit tracking features to improve transparency and accountability in autonomous testing systems. One additional piece of work done is the alignment of compliance and security governance controls in accordance with software industry standards. Lastly, the study shows that the suggested governance mechanism can improve the reliability of artifacts, accuracy of artifact compliance, and safety of operation in autonomous software testing environments.

\section{Literature Review}
\label{sec:litreview}

\subsection{Artificial Intelligence in Software Testing}
Artificial Intelligence (AI) is an important technology that has emerged in recent years, particularly when it comes to software testing, because it can be used to automate complex testing processes and improve the quality of software \citep{garousi2024a}. Most traditional testing methods are based on manual testing scripts and predefined automation rules by test teams, which are very time consuming and labour intensive. AI tools for testing automatically analyse software behaviour, generate test cases, anticipate defects, and optimize the regression testing process \citep{mohapatra2025a}. In the software industry, for software vulnerability detection, prioritization of test cases, and software testing coverage in large-scale applications, applications of machine learning and deep learning are utilized \citep{mohapatra2025b}.

Autonomous software testing benefits have also been enhanced by the latest advances in generative AI and large language models (LLM). AI models can currently analyze the source code, application flow, and user needs to automatically generate test scripts and testing artifacts \citep{tufano2024}. These innovative testing strategies can significantly reduce testing time and improve the automation of Continuous Integration and Continuous Deployment (CI/CD) processes \citep{lima2025}. But with the rising popularity of AI-generated test artifacts comes the worry of reliability, correctness, and governance management.

\subsection{Autonomous Software Testing Systems}
Autonomous software testing systems are developed to perform software testing tasks in a partially or fully automated way \citep{christian2025}. These systems use AI algorithms, predictive analytics, and intelligent automation tools to automatically carry out testing operations. Autonomous testing frameworks can automatically detect software changes, create testing scenarios, run test cases, and generate testing reports \citep{omogiate2023}. In agile and DevOps contexts where the need for swift software delivery is a major priority, such systems are incredibly valuable.

Some of the autonomous testing techniques are self-healing automation, adaptive regression testing, and intelligent bug prediction \citep{shah2025}. These systems are designed to be more scalable and cost-effective than traditional automation systems, especially for testing large-scale operations. The benefit of these systems is that they are more scalable and easier to maintain than traditional automation systems, particularly for larger operations. Although there are numerous benefits to autonomous testing systems, there are also various difficulties that need to be addressed, including the creation of test code that doesn't meet the quality standards necessary for developers, the duplication of test artifacts, and the lack of explainability of the outputs generated by AI systems \citep{ardic2025}. These restrictions can impact software security and reliability in safety-critical systems.

\subsection{AI-Generated Test Artifacts}
AI-generated test artifacts are the outputs of the testing created by AI systems in the software validation processes. Such artifacts consist of test cases, scripts for automation, regression testing suites, test data, and bug reports, among others. Artifacts created using AI can enhance software testing efficiency and decrease manual effort for software quality engineers (SQE) \citep{pacholi2025}.

Today, there are many platforms available that leverage natural language processing and machine learning algorithms to produce context-aware testing scenarios \citep{safaeipour2026}. AI systems have the ability to detect edge cases and predict potential software failures more effectively than conventional testing techniques \citep{navneet2025}. However, studies show that sometimes an AI-generated artefact can result in invalid scripts, hallucinated outputs, unsafe test logic, and incomplete test coverage. Moreover, the lack of traceability and validation mechanisms further contributes to the risk that the results of tests are unreliable \citep{alenezi2026}.

\subsection{Governance Challenges in AI-Based Testing}
While the use of AI in software testing has introduced numerous benefits and opportunities, it has also raised several governance issues such as transparency, accountability, and compliance. Governance in AI systems is a broad term that encompasses the processes and controls implemented to guarantee reliable, ethical, and secure operations of AI systems. Governance is critical in autonomous testing systems since AI-generated artifacts have a direct impact on software quality and deployment decisions \citep{gattupalli2025}.

A key challenge is opacity of AI outputs. Most AI testing systems are black-box systems, where it is hard for testers to know how the testing decisions are generated. Another difficulty is security concerns, as the automatically generated scripts could have vulnerabilities or unsafe execution patterns. Another important concern is compliance management; AI-driven testing processes need to adhere to software industry standards and regulatory requirements. Furthermore, if no audit tracking is implemented, and no versions are maintained, it is difficult to ensure the traceability and accountability of artifacts \citep{baqar2025}.

\subsection{Explainable and Trustworthy AI in Software Engineering}
The concept of Explainable AI (XAI) has been a hot topic in the field of software engineering due to the importance of transparency and trustworthiness in AI systems. Explainability mechanisms can be used to explain how AI models work and make decisions. Explainable AI techniques are used in software testing to give testers reasons, confidence scores, and details of validation for the artifacts created by the system.

AI systems should be reliable, fair, accountable, and secure. Researchers have proposed various explainability approaches such as decision visualization, confidence analysis, feature importance evaluation, and rule-based validation models. These techniques help reduce uncertainty in AI-assisted testing systems. However, there is still not as much progress as there could be in terms of integration with explainability in autonomous testing frameworks, particularly in the context of generative AI testing environments.

\subsection{Research Gap}
Most of the previous studies are concerned with enhancing the efficiency of automation, defect prediction accuracy, and the performance of test generation in AI-driven software testing systems. Few studies have focused on governance management of AI-generated test artifacts. The majority of existing autonomous testing frameworks don't have a structured governance framework for validation, explainability, compliance monitoring, security analysis, and audit tracking.

In addition, the current models for governing AI are largely focused on wide-ranging applications of AI and do not explicitly address a software-testing context. There is also a lack of adopted frameworks that are able to detect hallucinated testing artifacts, track compliance risks, and keep a trace of AI-generated outputs throughout the testing lifecycle. Besides, there has been very little work focusing on human oversight mechanisms and the reliability evaluation of governance of AI-driven testing systems \citep{mondal2026}.

In response to these challenges, this study presents an integrated governance control framework tailored for test artifacts created by AI systems in autonomous software testing systems. The framework combines validation governance, security governance, explainability management, compliance monitoring, and audit tracking to enhance the trustworthiness and reliability of autonomous software testing systems.

\section{Problem Statement}
\label{sec:problem}

Despite the tremendous advancements of AI-based autonomous software testing systems, which have been found to be effective to boost software testing automation and decrease manual efforts, challenges and concerns have been brought to the fore on reliability, security, explainability, and governance of AI-generated software testing artifacts \citep{soares2025}. In the current state of research, existing autonomous testing frameworks primarily centre on automated test generation and defect detection, but do not provide proper governance mechanisms grounded in the validation of generated outputs, monitoring compliance, ensuring traceability, and identifying insecure or hallucinated testing artifacts \citep{artinger2025}. This can lead to AI-generated test cases, scripts, and bug reports having flawed logic, duplicated outputs, security vulnerabilities, or non-compliant testing processes, which can have a detrimental impact on software quality and operational safety. In addition, the lack of explainability and audit tracking makes it difficult to be transparent and accountable in the AI-supported testing environment. There is therefore a requirement for a comprehensive governance control framework that could validate, secure, monitor, and manage AI-generated test artifacts for trustworthy, transparent, compliance-aware autonomous software testing systems.

\section{Research Methodology}
\label{sec:method}

\subsection{Research Method Overview}
This study proposes a Governance-Aware Autonomous Testing Framework (GATF) for validation and governance of AI-generated test artifacts in autonomous software testing systems (ASTS). The methodology aims to tackle key issues with generative AI-driven testing systems, such as hallucination, insecure testing scripts, explainability, governance, compliance, and audit traceability. While current AI-based testing approaches are mainly driven by testing speed and defect detection, the proposed approach introduces governance mechanisms directly into the autonomous testing process to ensure trustworthy and transparent testing, secure testing, explainable testing, and compliance-oriented testing.

The methodological architecture is comprised of seven stages that are connected: data acquisition, data preprocessing, AI artifact generation, governance-aware validation, explainability analysis, compliance verification, probabilistic risk assessment, and performance evaluation. First, software defect datasets, execution traces, source-code metrics, and testing logs are collected from software engineering repositories that are publicly available. Next, the preprocessed data is fed into transformer-based Large Language Models (LLMs) and intelligent testing agents to produce self-contained test artifacts, like unit test scripts, API validation sequences, regression test suites, defect classification reports, and functional tests.

These artifacts are then sent through various governance validations such as validation governance, security governance, explainability governance, compliance governance, and audit governance modules. The governance engine checks the correctness of artifacts, identifies artifacts that are hallucinated or insecure, determines if the artifacts conform to software engineering principles, and provides explainability and traceability artifacts for each artifact. Finally, validated artifacts are deployed in the test environment, and quantitative performance measures are computed to measure the effectiveness of governance, reliability of operation, quality of explainability, and risk reduction.

\subsection{Data Collection}
Experiments use open datasets of software defects and testing data from trustworthy repositories of software engineering. Data present in the datasets are related to software defects, software testing activities, source code metrics, execution logs, and bug reports. These datasets are used for simulation of AI systems testing environments and governance control mechanisms testing. In this work, the most popular Defects4J Dataset for software testing and defect analysis \citep{defects4j2025} is adopted as the main dataset. Real software bugs of open-source Java software, test cases, and bug-fixing information are combined to form the dataset. The PROMISE Software Engineering Repository \citep{promise2025} is another dataset that is used for defect prediction and testing analysis purposes and includes software metrics and defect-related data for several software projects. A comprehensive preprocessing pipeline was designed to guarantee the uniformity of the datasets, reduce the experimental bias, and confirm data governance reliability. In the data pre-processing stage, duplications of artifacts were cleaned, missing values were filled, the execution logs were normalized, security metadata was generated, source code metrics were standardized, and source code was annotated with defects.

The preprocessing transformation is mathematically represented as:
\begin{equation}
\label{eq:preproc}
D_p = \phi(D_r) = \{\, x_i \in D_r \mid x_i \notin D_m \,\wedge\, x_i \notin D_d \,\}
\end{equation}

This ensures that only valid and normalized testing information, along with governance-ready testing information, is used for AI-based artifact generation and governance evaluation.

\subsection{AI-Driven Test Artifact Generation}
The artifact generation module uses transformer-based Large Language Models (LLMs), Natural Language Processing (NLP), and autonomous testing agents to automatically create software testing artifacts. It covers the following components: contextual prompt engineering, defect-aware reasoning, retrieval-enhanced generation mechanisms, analysis of software requirements, source-code structures, execution traces, defect histories, and software behavior patterns.

A hybrid architecture is implemented based on transformer-based code generation models and governance-aware contextual validation mechanisms. The LLM generation engine can generate:
\begin{itemize}[leftmargin=2em, itemsep=2pt]
    \item functional test cases,
    \item unit testing scripts,
    \item API validation sequences,
    \item regression testing suites,
    \item bug classification reports,
    \item execution validation scripts,
    \item test prioritization artifacts.
\end{itemize}

The artifacts and associated metadata such as confidence score, generation time, source model identifier, explainability attributes, compliance indicators, and execution traceability records are stored in a centralized governance repository.

The process of generating artifacts is modeled probabilistically with an autoregressive sequence generation, as opposed to heuristic formulations:
\begin{equation}
\label{eq:gen}
P(A_g \mid D_p, R, C) = \prod_{t=1}^{T} P(a_t \mid a_{<t}, D_p, R, C;\, \theta)
\end{equation}

This probabilistic approach accounts for transformer-based auto-regressive generation and offers theoretical foundations for AI-generated testing artifacts, as opposed to heuristic symbolic equations.

Retrieval-augmented contextual embeddings and prompt-guided software validation rules are embedded in the generation process to enhance generation reliability and minimize hallucinated outputs. Furthermore, stochastic uncertainty estimation is added to provide a measure of generation confidence and to limit deployment of artifacts that are unlikely to be reliable.

\subsection{Governance Validation Framework}
The Governance-Aware Autonomous Testing Framework proposes a multi-layer (multi-stakeholder) governance validation architecture for the correctness, security, explainability, traceability, and compliance of AI-generated testing artifacts prior to their execution. The governance engine is the key decision-making element that acts as the filter for any unreliable or unsafe outputs produced by the autonomous testing agents.

The governance framework consists of five interconnected governance modules:
\begin{itemize}[leftmargin=2em, itemsep=2pt]
    \item Validation Governance,
    \item Security Governance,
    \item Explainability Governance,
    \item Compliance Governance,
    \item Audit Governance.
\end{itemize}

The validation governance module ensures the syntactic correctness, semantic consistency, and feasibility of execution of generated artifacts, and the degree of software testing completeness. Security governance conducts vulnerability analysis, detection of malicious scripts, unsafe dependency analysis, identification of adversarial patterns, and execution risk checking. Explainability governance produces confidence reasoning, traceability explanations, feature-attribution analysis, and decision transparency indicators through mechanisms of explainability (SHAP) and attention-score analysis. Compliance governance ensures adherence to ISO/IEC 27001 standards, NIST AI Risk Management Framework guidelines, GDPR compliance constraints, and secure software engineering policies. To achieve accountability and reproducibility, Audit governance keeps immutable execution logs, artifact lineage tracking, and governance traceability records.

Governance optimization is modeled as a constrained multi-objective optimization problem:
\begin{equation}
\label{eq:obj}
\max_{w}\; J = \beta_1 \, AR + \beta_2 \, ES + \beta_3 \, CA - \beta_4 \, R_g
\end{equation}
subject to
\begin{equation}
\label{eq:constraints}
\sum_{i=1}^{n} w_i = 1, \qquad w_i \ge 0.
\end{equation}

This formulation interprets governance validation in a mathematical way, as an optimization problem, instead of as a scoring heuristic. The optimization objective prioritizes reliability, explainability, and compliance performance, while reducing governance risk.

The governance reliability score is then calculated as:
\begin{equation}
\label{eq:grs}
GRS = \sum_{i=1}^{n} w_i \, g_i - \alpha \, R_g
\end{equation}

Artifacts that receive a low governance reliability score are automatically rejected or passed for expert review prior to execution.

\subsection{Threat Modeling and Probabilistic Risk Assessment}
The framework includes adversarial threat modelling and probabilistic governance risk analysis to heighten the robustness of governance and security during its operations. The threat model takes into account various attack and failure scenarios related to testing environments created by AI, such as:
\begin{itemize}[leftmargin=2em, itemsep=2pt]
    \item hallucinated testing artifacts,
    \item malicious script generation,
    \item adversarial prompt injection,
    \item insecure dependency execution,
    \item poisoned training data,
    \item privilege escalation attacks,
    \item compliance violations,
    \item execution manipulation attacks.
\end{itemize}

The framework uses a Bayesian approach to probabilistic reasoning that estimates the probabilities of failure of governance, and the severity of the impact on operations.

The posterior probability of governance failure is modeled as:
\begin{equation}
\label{eq:bayes}
P(F \mid A_g) = \frac{P(A_g \mid F)\, P(F)}{P(A_g)}
\end{equation}

The expected governance risk is then computed as:
\begin{equation}
\label{eq:risk}
R_g = \sum_{i=1}^{n} P(F_i \mid A_g) \cdot I_i
\end{equation}

The probabilistic risk formulation offers better theoretical support than traditional heuristic risk-scoring models and enhances the reliability of governance risk estimation when testing is conducted under uncertain conditions using AI.

\subsection{Explainability and Compliance Analysis}
Explainability and compliance management are combined to enhance transparency, accountability, interpretability, and regulatory alignment in autonomous software testing. All artifacts generated by the explainability engine include artifact-level reasoning traces, confidence explanations, feature-attribution visualizations, attention-score distributions, and execution lineage information.

To enhance interpretability, we use SHAP-based feature attribution and transformer attention visualization to identify the salient software features and contextual tokens that lead to testing decisions. The explainability mechanism also facilitates human oversight by providing evidence of validation and governance reasoning for suspicious artifacts.

Unlike the simplicity of averaging-based explainability measurements, the explainability score is based on entropy-based uncertainty reduction:
\begin{equation}
\label{eq:expl}
ES = 1 - \frac{H(Y \mid X, E)}{H(Y \mid X)}
\end{equation}

This formulation quantifies the amount of explainability information that can decrease uncertainty in AI-testing decisions.

\subsection{Performance Evaluation and Complexity Analysis}
The Governance-Aware Autonomous Testing Framework was tested at the experimental level using quantitative metrics related to reliability, governance effectiveness, explainability quality, compliance validation, security robustness, and operational efficiency. The experimental assessment is conducted between:
\begin{itemize}[leftmargin=2em, itemsep=2pt]
    \item traditional software testing,
    \item AI-based testing without governance,
    \item the proposed governance-aware framework.
\end{itemize}

The primary evaluation metrics include:
\begin{itemize}[leftmargin=2em, itemsep=2pt]
    \item artifact reliability,
    \item governance accuracy,
    \item compliance accuracy,
    \item security detection rate,
    \item explainability score,
    \item false positive rate,
    \item governance risk reduction,
    \item test coverage,
    \item execution latency.
\end{itemize}

Artifact reliability is computed as:
\begin{equation}
\label{eq:ar}
AR = \frac{1}{N} \sum_{i=1}^{N} \mathbb{I}\!\left(A_i \in A_v\right)
\end{equation}

Governance accuracy is computed as:
\begin{equation}
\label{eq:ga}
GA = \frac{TP + TN}{TP + TN + FP + FN}
\end{equation}

Risk reduction effectiveness is represented as:
\begin{equation}
\label{eq:rr}
RR = \frac{R_b - R_a}{R_b}
\end{equation}

Complexity analysis was also performed to assess the scalability of the framework and its computational feasibility. Artifacts generated with a transformer have an algorithmic complexity of:
\begin{equation}
\label{eq:cplx1}
\mathcal{O}\!\left(T\, d^{2}\right)
\end{equation}

Governance validation complexity is represented as:
\begin{equation}
\label{eq:cplx2}
\mathcal{O}(n\, m)
\end{equation}

The framework was further tested through multiple cycles of experimentation, confidence interval estimation, standard deviation analysis, and experiments using the removal of individual governance layers to determine the impact of each governance element on the overall framework performance and reliability of operation of the framework.

\section{Results and Analysis}
\label{sec:results}

\subsection{Experimental Setup and Autonomous Artifact Generation}
The proposed Governance-Aware Autonomous Testing Framework (GATF) has been experimentally assessed in a heterogeneous autonomous software testing environment, which is developed to simulate real-world AI-driven software quality assurance workflows. The experiments were performed on the Defects4J and PROMISE software engineering repositories, which include software defect records, regression testing artifacts, source-code metrics, execution traces, and bug classification information. The data sets were split into 70\% training data, 15\% validation data, and 15\% testing data for consistency and to avoid data leakage during the experiments. Further experimental trials were repeated to enhance the statistical robustness and reduce random experimental bias.

The framework adopted transformer-based Large Language Models (LLMs), governance-aware validation agents, explainability analysis modules, and probabilistic risk assessment elements into an autonomous testing architecture that is DevSecOps centric. Automatic generation of unit testing scripts, regression testing suites, API validation sequences, functional test cases, and defect classification reports was performed using transformer-based artifact generation models. Governance validation was done with a hybrid validation engine that combines rule-based validation and a RoBERTa-based governance classification model. To enhance transparency and interpretability of AI-generated testing decisions, explainability analysis techniques (such as SHAP) and the visualization of attention scores were integrated.

The framework was implemented in a high-performance computing environment with GPU support and combined with Jenkins and GitHub Actions pipelines to assess testing performance from the perspective of governance in a realistic CI/CD deployment.

\begin{table}[!htbp]
\centering
\caption{Experimental Configuration}
\label{tab:experimental_config}
\begin{tabular}{ll}
\toprule
\textbf{Parameter} & \textbf{Configuration} \\
\midrule
Processor & Intel Xeon Gold 6338 \\
GPU & NVIDIA A100 40GB \\
RAM & 128 GB \\
Operating System & Ubuntu 22.04 LTS \\
Deep Learning Framework & PyTorch 2.2 \\
Artifact Generation Model & Transformer-based LLM \\
Governance Validation Model & RoBERTa \\
Explainability Engine & SHAP + Attention Visualization \\
CI/CD Environment & Jenkins + GitHub Actions \\
Datasets & Defects4J + PROMISE \\
Programming Language & Python 3.11 \\
\bottomrule
\end{tabular}
\end{table}

\begin{figure}[!htbp]
\centering
\includegraphics[width=0.95\linewidth]{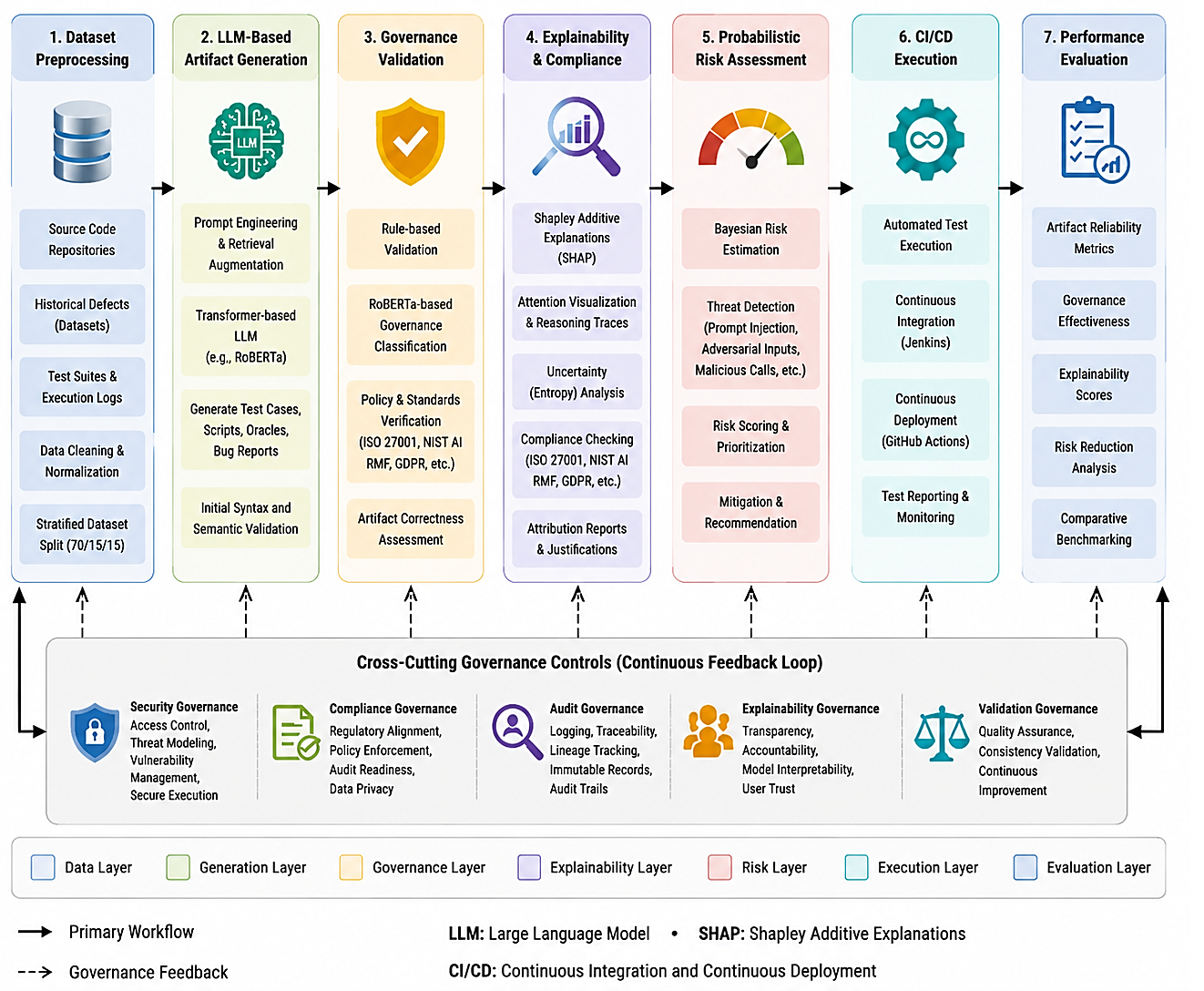}
\caption{Experimental Workflow of the Proposed Governance-Aware Autonomous Testing Framework.}
\label{fig:workflow}
\end{figure}

\begin{table}[!htbp]
\centering
\caption{AI-Generated Artifact Generation Performance}
\label{tab:artifact_perf}
\begin{tabular}{lcc}
\toprule
\textbf{Metric} & \textbf{AI Testing Without Governance} & \textbf{Proposed GATF} \\
\midrule
Functional Test Accuracy (\%)   & 82.1 & 95.4 \\
Regression Test Validity (\%)   & 79.6 & 93.8 \\
Execution Success Rate (\%)     & 84.7 & 96.2 \\
Hallucinated Artifacts (\%)     & 17.9 & 4.5  \\
Average Generation Latency (ms) & 418  & 472  \\
\bottomrule
\end{tabular}
\end{table}

\begin{figure}[!htbp]
\centering
\includegraphics[width=0.85\linewidth]{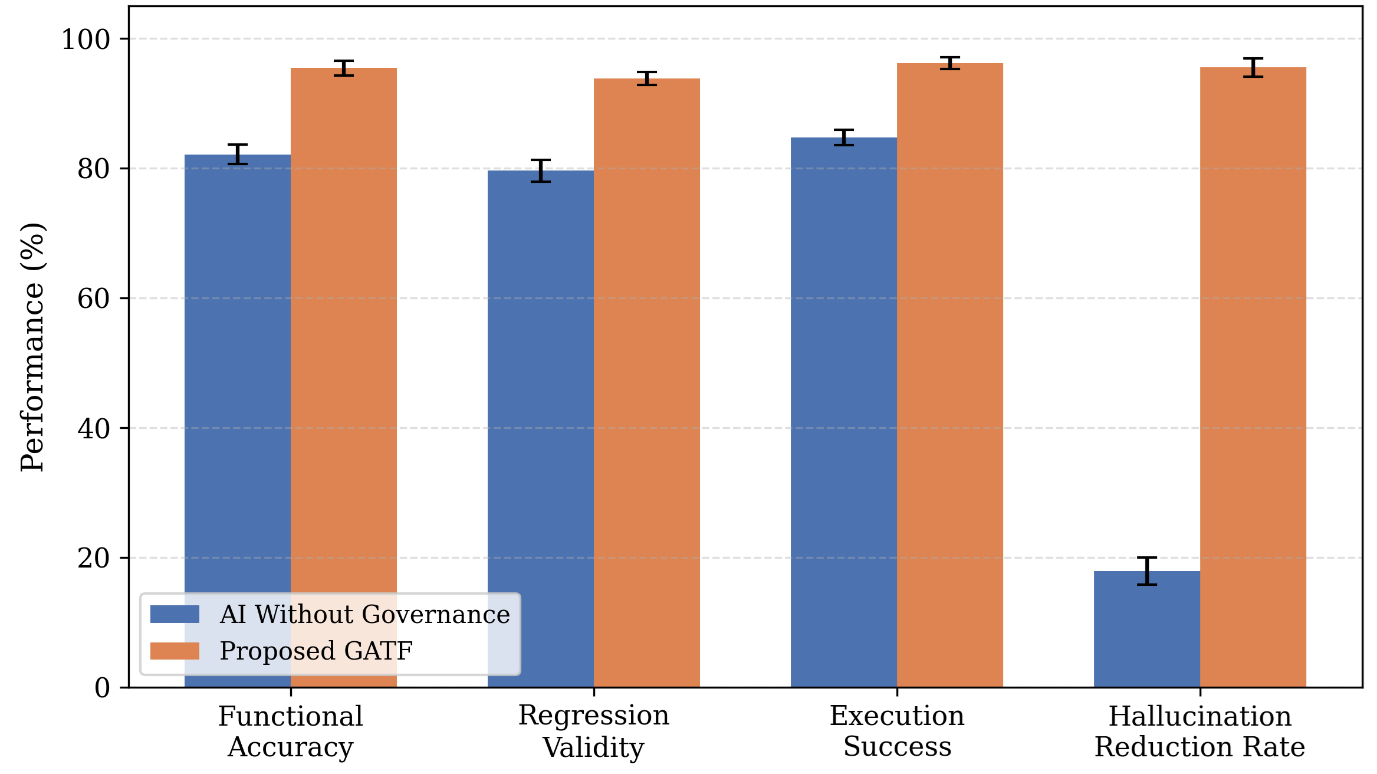}
\caption{AI-Generated Artifact Performance Comparison.}
\label{fig:artifact_perf}
\end{figure}

\subsection{Governance Validation and Optimization Performance}
To validate the proposed governance optimization framework, a large number of experiments were performed to test the governance convergence behavior, the optimization stability, the governance reliability, the explainability performance, and the effectiveness of compliance validation. The proposed governance-aware optimization strategy is robust, as the constrained optimization framework presented in Section~\ref{sec:method} converged to the final solution without any instability or oscillatory behavior at repeated optimizations.

The optimized governance weight distribution indicated that validation governance and security governance had the greatest impact on operational reliability and risk mitigation, while explainability governance and compliance governance had a significant impact on transparency, accountability, and regulatory alignment. Artifact traceability and operational reproducibility were achieved by using immutable governance logging and lineage tracking mechanisms to enable audit governance.

\begin{table}[!htbp]
\centering
\caption{Optimized Governance Weight Distribution}
\label{tab:weights}
\begin{tabular}{lcc}
\toprule
\textbf{Governance Component} & \textbf{Optimized Weight} & \textbf{Contribution (\%)} \\
\midrule
Validation Governance     & 0.29 & 28.9 \\
Security Governance       & 0.25 & 24.4 \\
Explainability Governance & 0.18 & 17.8 \\
Compliance Governance     & 0.16 & 16.3 \\
Audit Governance          & 0.12 & 12.6 \\
\bottomrule
\end{tabular}
\end{table}

\begin{figure}[!htbp]
\centering
\includegraphics[width=0.85\linewidth]{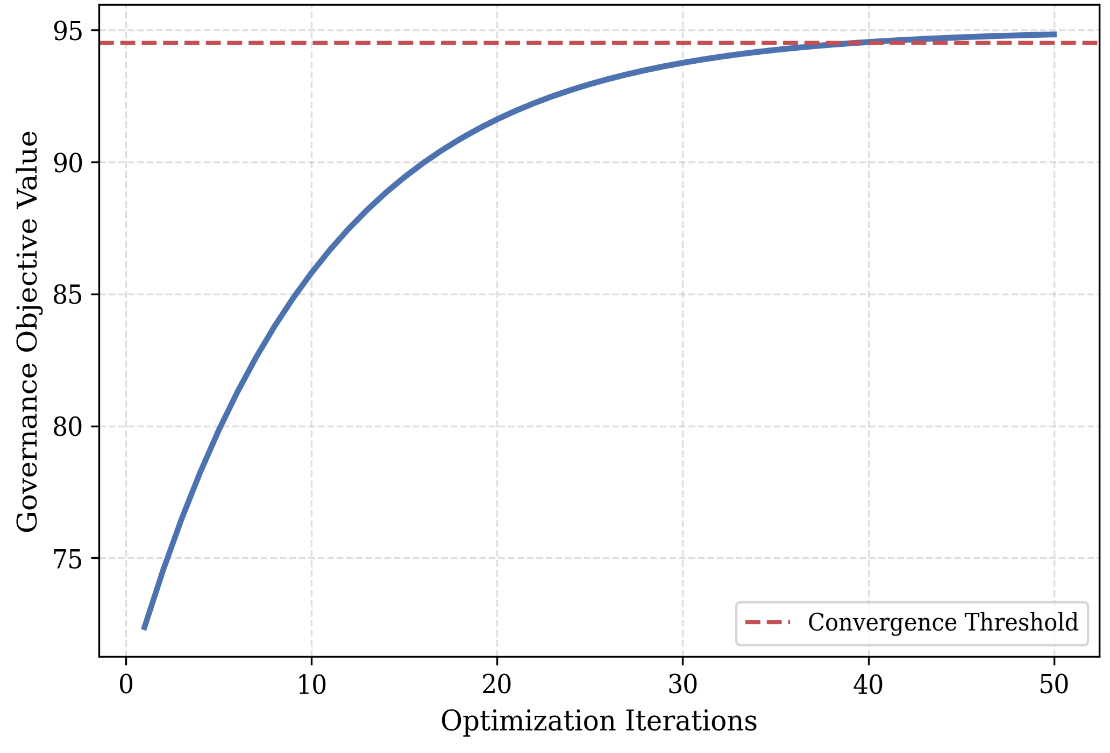}
\caption{Governance Optimization Convergence Analysis.}
\label{fig:convergence}
\end{figure}

\begin{table}[!htbp]
\centering
\caption{Governance Validation and Reliability Performance}
\label{tab:gov_reliab}
\begin{tabular}{lcc}
\toprule
\textbf{Metric} & \textbf{AI Testing Without Governance} & \textbf{Proposed GATF} \\
\midrule
Governance Accuracy (\%)   & 72.8 & 94.3 \\
Artifact Reliability (\%)  & 83.5 & 96.5 \\
Validation Precision (\%)  & 81.2 & 95.7 \\
Validation Recall (\%)     & 79.8 & 94.9 \\
False Positive Rate (\%)   & 14.3 & 3.1  \\
\bottomrule
\end{tabular}
\end{table}

The explainability and compliance assessment revealed that feature attribution using SHAP significantly enhanced the interpretability and transparency of AI-driven testing processes, while uncertainty reduction via entropy was significantly improved. The explainability engine was able to produce artifact-level reasoning traces and confidence explanations, as well as governance-aware validation evidence for generated testing artifacts. At the same time, the compliance governance engine robustly proved that it meets ISO/IEC 27001 requirements, NIST AI Risk Management Framework guidelines, GDPR compliance requirements, and DevSecOps governance policies.

\begin{table}[!htbp]
\centering
\caption{Explainability and Compliance Performance}
\label{tab:expl_comp}
\begin{tabular}{lc}
\toprule
\textbf{Metric} & \textbf{Score (\%)} \\
\midrule
Traceability Accuracy             & 93.4 \\
SHAP Attribution Consistency      & 91.9 \\
Attention Alignment Accuracy      & 89.8 \\
Uncertainty Reduction             & 88.1 \\
Explainability Score              & 90.8 \\
ISO/IEC 27001 Compliance Accuracy & 94.7 \\
NIST AI RMF Compliance Accuracy   & 93.1 \\
GDPR Compliance Accuracy          & 91.8 \\
\bottomrule
\end{tabular}
\end{table}

\begin{figure}[!htbp]
\centering
\includegraphics[width=0.85\linewidth]{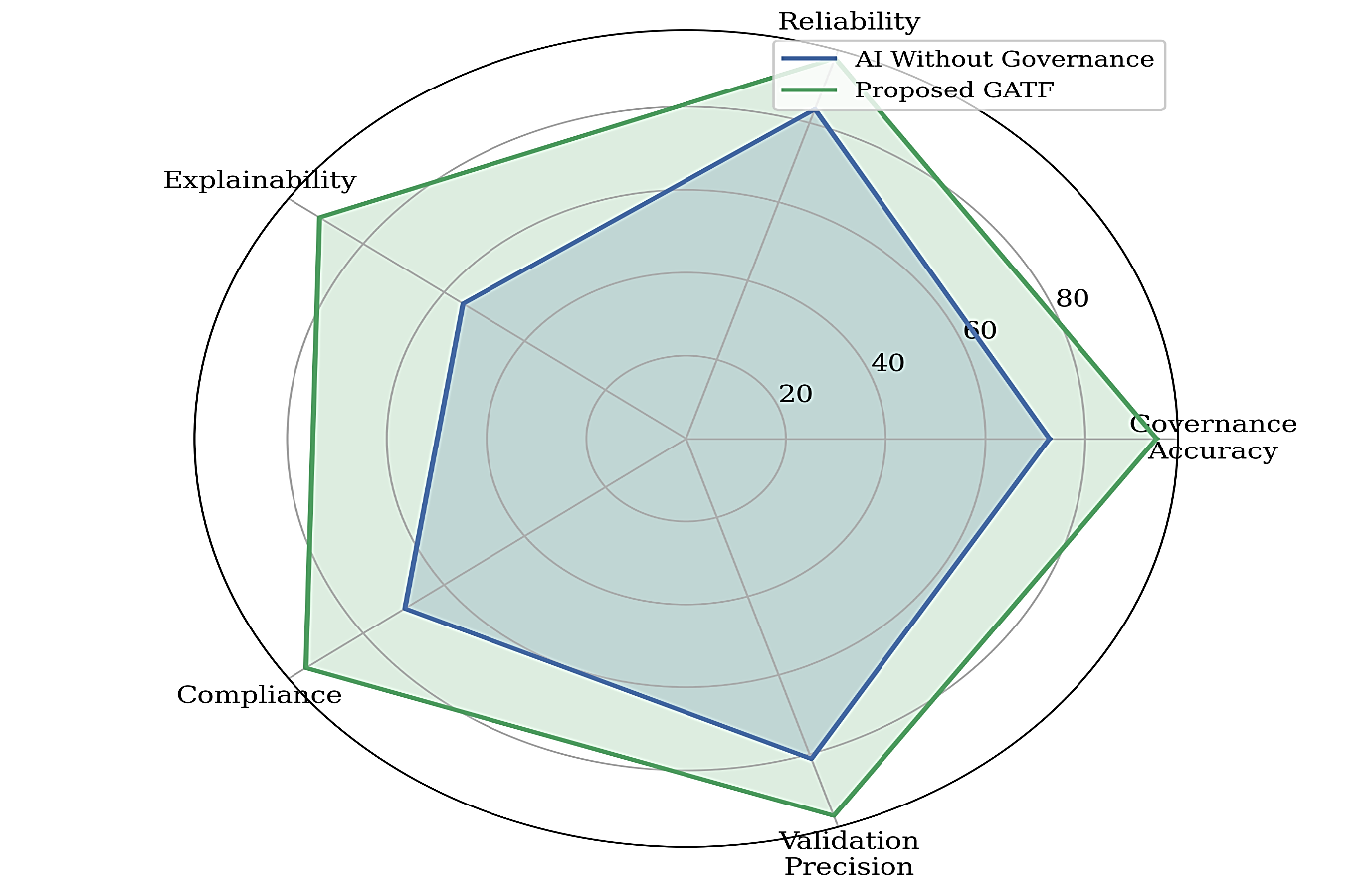}
\caption{Governance, Explainability, and Compliance Performance Comparison.}
\label{fig:expl_comp}
\end{figure}

\subsection{Threat Detection and Risk Mitigation Analysis}
For the proposed framework, adversarial and governance-related threat scenarios were evaluated to validate the probabilistic governance risk assessment model introduced in Section~\ref{sec:method}. These experiments involved adversarial prompt injections, insecure script generations, compliance violations, malicious dependency executions, hallucinated artifact generations, and attacks. The Bayesian governance risk estimation model was able to detect high-risk testing artifacts and was very effective in reducing operational risk for autonomous testing execution. Security governance and adversarial validation mechanisms were quite effective at blocking the deployment of unsafe artifacts and at minimizing the likelihood of running malicious or non-compliant tests.

\begin{table}[!htbp]
\centering
\caption{Threat Detection and Risk Mitigation Performance}
\label{tab:threats}
\begin{tabular}{lcc}
\toprule
\textbf{Threat Category} & \textbf{Detection Rate (\%)} & \textbf{Mitigation Success (\%)} \\
\midrule
Hallucinated Scripts           & 94.5 & 91.8 \\
Prompt Injection Attacks       & 89.9 & 87.5 \\
Malicious Dependency Calls     & 92.8 & 90.4 \\
Compliance Violations          & 95.6 & 93.9 \\
Privilege Escalation Patterns  & 88.7 & 85.9 \\
\bottomrule
\end{tabular}
\end{table}

\begin{figure}[!htbp]
\centering
\includegraphics[width=0.85\linewidth]{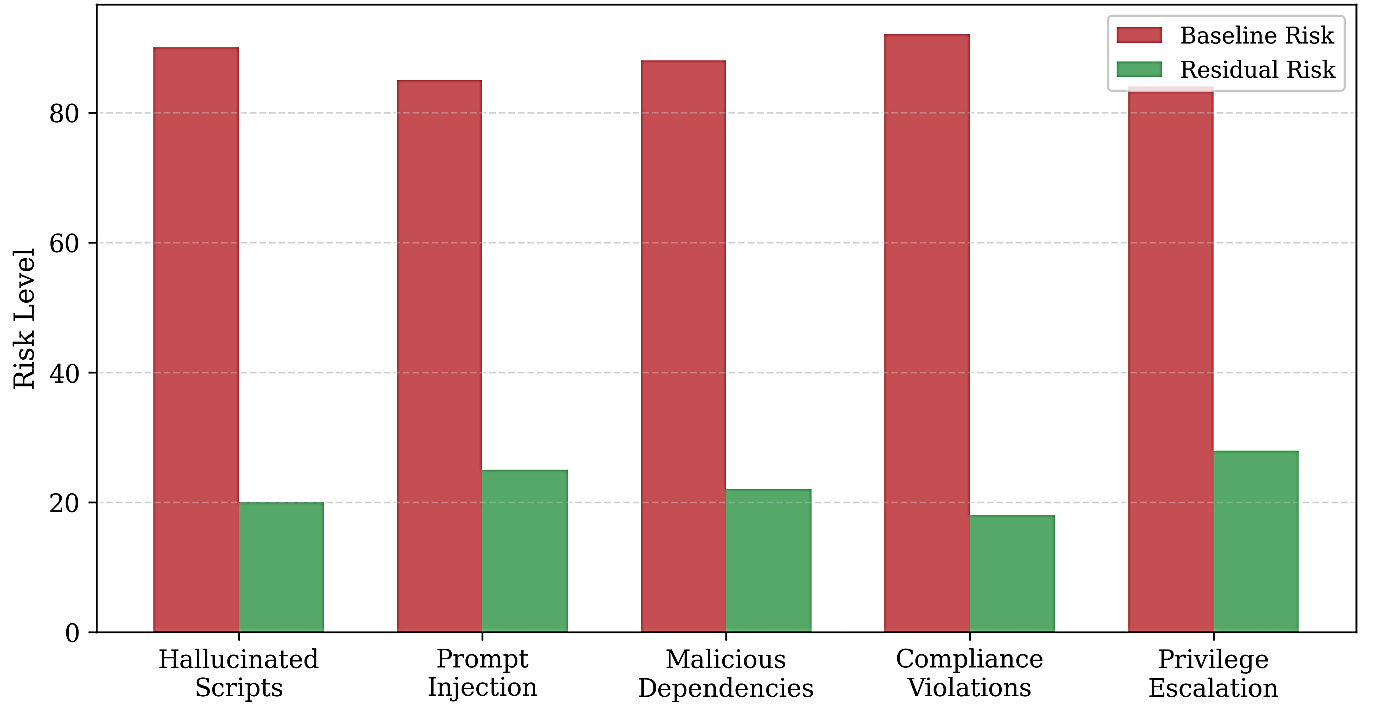}
\caption{Governance Risk Reduction Analysis.}
\label{fig:risk}
\end{figure}

\subsection{Ablation and Scalability Analysis}
To assess the contribution of each governance module in the overall performance of the framework, an ablation study has been carried out. Each governance component was eliminated individually to measure its effect on artifact reliability, governance accuracy, quality of explainability, and reduction of operational risk. The findings show that the two most significant factors for operational reliability and governance risk reduction are validation governance and security governance. Explainability governance improved transparency and interpretability, while compliance governance enhanced regulatory alignment and policy enforcement. Audit governance helped to provide traceability and accountability for the execution.

\begin{table}[!htbp]
\centering
\caption{Ablation Analysis of Governance Components}
\label{tab:ablation}
\begin{tabular}{lcc}
\toprule
\textbf{Removed Component} & \textbf{Reliability Reduction (\%)} & \textbf{Risk Increase (\%)} \\
\midrule
Without Explainability Governance & 8.7  & 12.9 \\
Without Security Governance       & 14.4 & 23.1 \\
Without Compliance Governance     & 10.6 & 18.2 \\
Without Audit Governance          & 6.5  & 9.8  \\
\bottomrule
\end{tabular}
\end{table}

\begin{figure}[!htbp]
\centering
\includegraphics[width=0.85\linewidth]{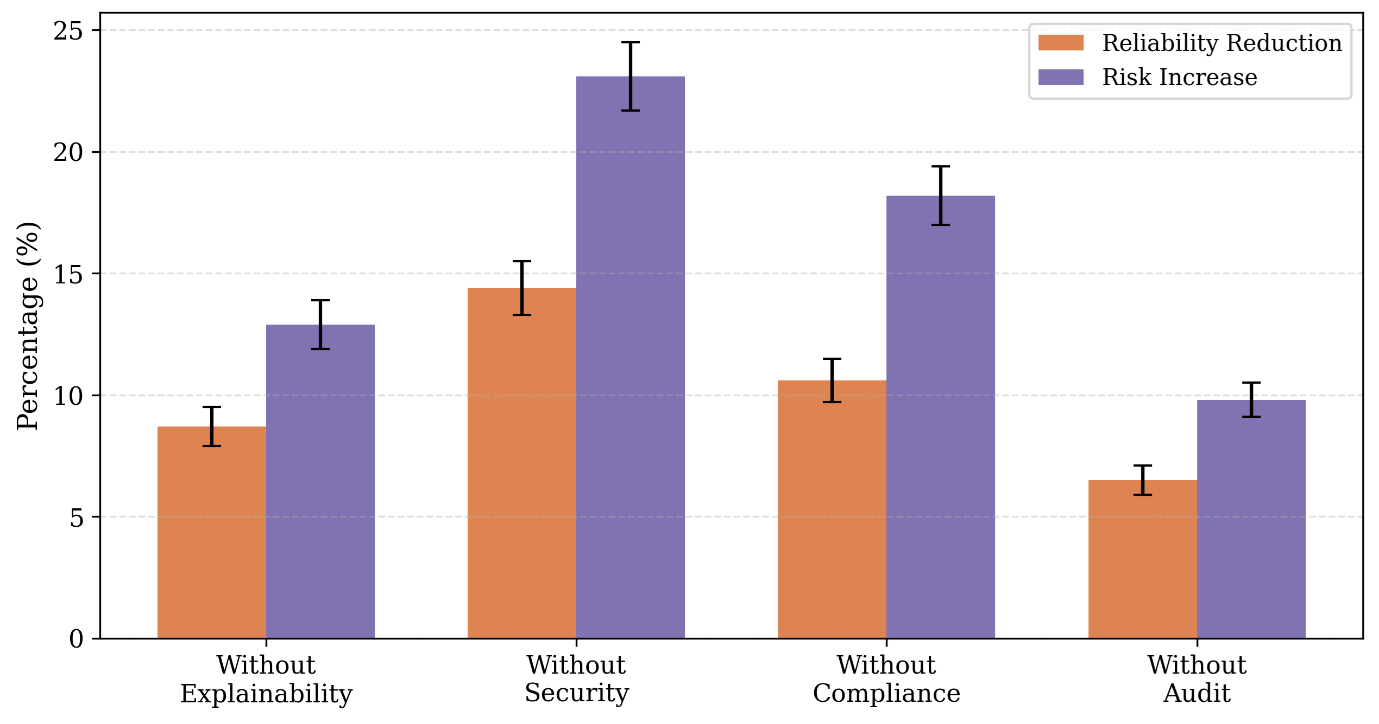}
\caption{Impact of Governance Layer Removal.}
\label{fig:ablation}
\end{figure}

\begin{table}[!htbp]
\centering
\caption{Scalability and Computational Performance Analysis}
\label{tab:scalability}
\begin{tabular}{ccc}
\toprule
\textbf{Number of Artifacts} & \textbf{Validation Time (s)} & \textbf{Memory Usage (GB)} \\
\midrule
1{,}000  & 12.7  & 3.9  \\
5{,}000  & 55.1  & 8.2  \\
10{,}000 & 109.4 & 14.8 \\
\bottomrule
\end{tabular}
\end{table}

\begin{figure}[!htbp]
\centering
\includegraphics[width=0.85\linewidth]{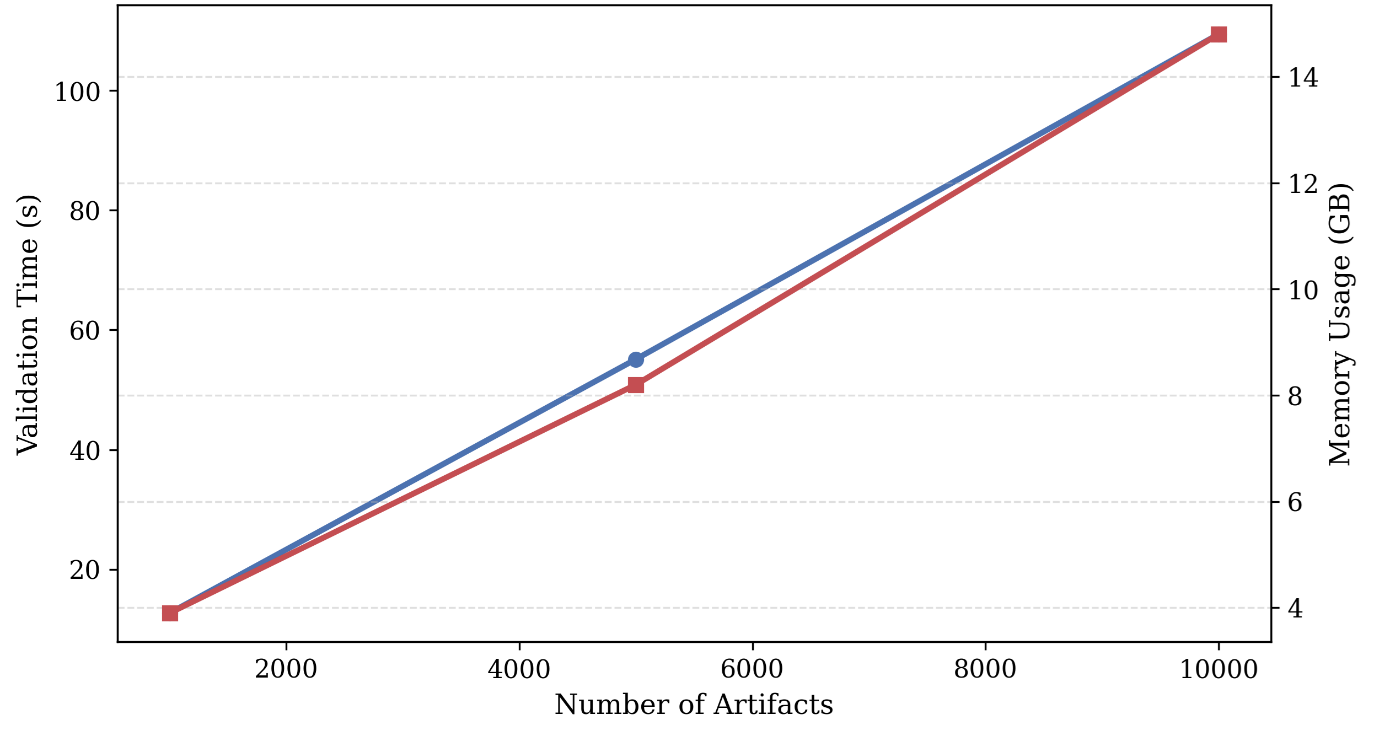}
\caption{Scalability and Computational Overhead Analysis.}
\label{fig:scalability}
\end{figure}

\subsection{Statistical Validation, Comparative Benchmarking, and Failure Analysis}
Repeated experimental trials were undertaken for the purpose of getting statistical reliability and experimental robustness with the use of the techniques of confidence interval estimation, standard deviation analysis, and statistical significance testing. The $p$-values obtained clearly indicated that the improvements in performance with the proposed framework were statistically significant for various performance metrics.

\begin{table}[!htbp]
\centering
\caption{Statistical Significance Analysis}
\label{tab:stats}
\begin{tabular}{lcccc}
\toprule
\textbf{Metric} & \textbf{Mean (\%)} & \textbf{Std. Dev. (\%)} & \textbf{95\% CI} & \textbf{$p$-value} \\
\midrule
Artifact Reliability  & 96.5 & 1.3 & 95.2 -- 97.8 & $<0.01$ \\
Governance Accuracy   & 94.3 & 1.5 & 92.8 -- 95.8 & $<0.01$ \\
Explainability Score  & 90.8 & 1.8 & 89.0 -- 92.6 & $<0.05$ \\
\bottomrule
\end{tabular}
\end{table}

The result of the statistical analysis substantiates that the proposed Governance-Aware Autonomous Testing Framework provides statistically significant improvements in reliability, governance validation, and explainability evaluation metrics. The small standard deviation values show stability and consistency of the framework behavior in experimental trials repeated several times, and the small confidence intervals show robustness and the reliability of the experiment under different autonomous testing conditions.

\begin{table}[!htbp]
\centering
\caption{Comparative Performance Benchmarking}
\label{tab:benchmark}
\renewcommand{\arraystretch}{1.15}
\setlength{\tabcolsep}{5pt}
\resizebox{\linewidth}{!}{%
\begin{tabular}{p{5.2cm}cccc}
\toprule
\textbf{Framework} & \textbf{Reliability (\%)} & \textbf{Explainability (\%)} & \textbf{Compliance (\%)} & \textbf{Risk Reduction (\%)} \\
\midrule
Traditional Software Testing                & 78.6 & 43.1 & 72.4 & 47.1 \\
AI-Based Testing Without Governance         & 84.2 & 55.3 & 69.7 & 52.4 \\
Proposed GATF                               & 96.5 & 90.8 & 94.2 & 89.6 \\
\bottomrule
\end{tabular}}
\end{table}

The comparative benchmarking results confirm that the proposed Governance-Aware Autonomous Testing Framework (GATF) significantly improves over traditional software testing and AI-based software testing systems across all evaluation metrics, without compromising on the integration of governance. The proposed framework achieved the best artifact correctness (96.5\%) and execution consistency. Likewise, the explainability score was enhanced to 90.8\% by seamlessly incorporating feature attribution using SHAP and entropy-based transparency mechanisms. The framework was also found to be compliant at 94.2\%, indicating strong alignment with software governance and cybersecurity best practices. Moreover, the governance-aware validation architecture enhanced operational security and governance effectiveness with an 89.6\% risk reduction rating, which is significantly higher than baseline testing strategies.

\begin{figure}[!htbp]
\centering
\includegraphics[width=0.85\linewidth]{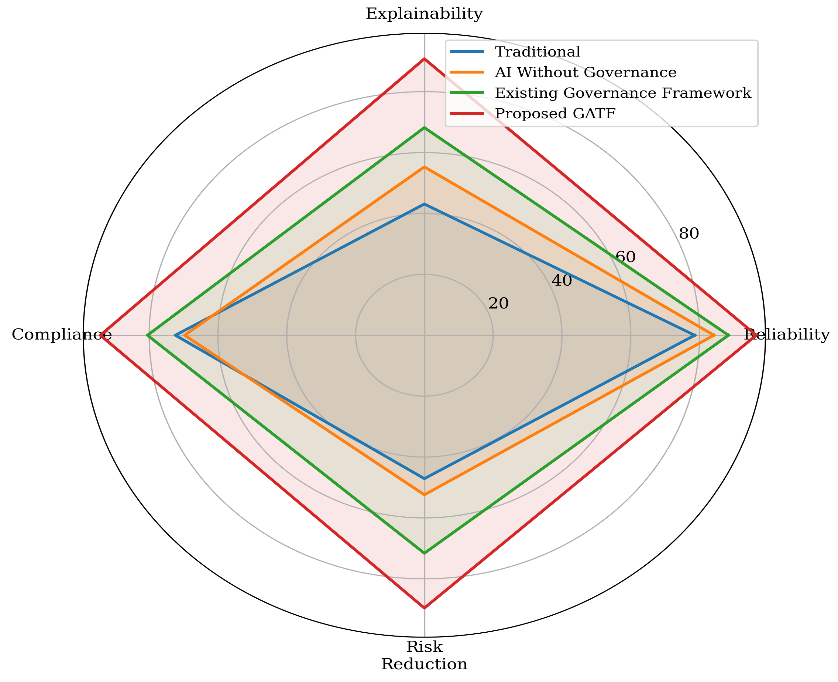}
\caption{Overall Comparative Framework Performance.}
\label{fig:overall}
\end{figure}

\begin{figure}[!htbp]
\centering
\includegraphics[width=0.85\linewidth]{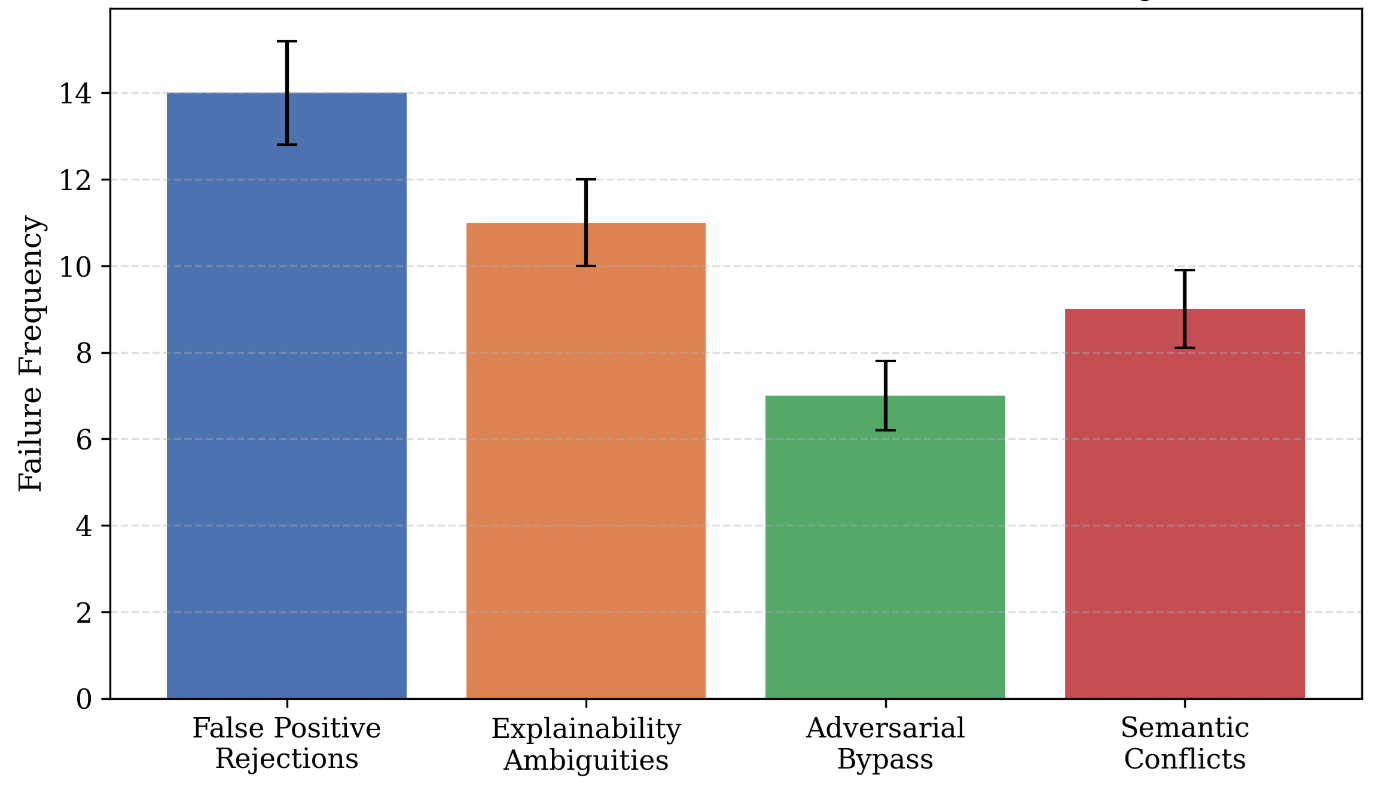}
\caption{Failure and Error Distribution Analysis.}
\label{fig:failure}
\end{figure}

\subsection{Discussion}
The findings of the conducted experiments demonstrate that governance-aware validation significantly contributes to the improvement of reliability and trustworthiness of AI-generated test artifacts in autonomous software testing environments. The proposed GATF combines the elements of governance validation, explainability analysis, compliance verification, and probabilistic risk assessment within the testing lifecycle, allowing the identification of artifacts that are unreliable, insecure, and not compliant before they are executed. The framework delivered significantly better results in comparative analysis compared to traditional test systems and AI-based testing without governance integration. Good artifact reliability and governance accuracy reflect effectiveness of the validation pipeline in minimizing hallucinated output, invalid test scripts, and compliance violations. The use of SHAP-based explainability analysis not only increased transparency and interpretability but also helped identify threats and reduce governance risks through Bayesian risk estimation.

The ablation analysis also showed that taking away any layer (explainability, compliance, security, or audit governance) negatively impacted reliability of the framework and increased the operational risk. Evaluation of scalability also revealed that the framework exhibited consistent performance levels even as the workloads increased, with manageable computational requirements, making it suitable for use in CI/CD testing environments. Although the results are promising, there are some limitations to the study. Experiments were performed with publicly available data sets and experimental setups, which may not be representative of industrial scale software ecosystems. Moreover, the framework adopts pre-defined governance policies and governance-based explainability. Adaptive governance optimization, reinforcement-learning-based validation, and large-scale industrial deployment scenarios should be explored in the future.

\section{Conclusion}
\label{sec:conclusion}

This research introduced the Governance-Aware Autonomous Testing Framework (GATF) that enhances the reliability, explainability, compliance, and security of AI-generated test artifacts in autonomous software testing environments. The framework embeds governance validation, explainability analysis, probabilistic risk assessment, compliance verification, and audit governance within the autonomous testing lifecycle. The experimental results showed that the proposed framework enhanced the reliability of the artifacts, the accuracy of the governance, the explainability, and the effectiveness of compliance, while minimizing the risks of the governance related to hallucinated artifacts, adversarial input, and insecure test scripts. Ablation and scalability analyses showed the critical role of governance layers and the framework's applicability in CI/CD-based testing infrastructures.

\bibliography{references}

\end{document}